# A spherical harmonics method for processing anisotropic X-ray atomic pair distribution functions

Authors

**Guanjie Zhang[a], Hui Liu[b], Jun Chen[b], He Lin[cd]\* and Nan Zhang[a]\***

[a]Electronic Materials Research Laboratory, Key Laboratory of the Ministry of Education and International Center for Dielectric Research, School of Electronic and Information Engineering, Xi'an Jiaotong University, Xi'an, 710049, People's Republic of China

[b]Beijing Advanced Innovation Center for Materials Genome Engineering, Department of Physical Chemistry, University of Science and Technology Beijing, Beijing, 100083, People's Republic of China

[c] Shanghai Synchrotron Radiation Facility, Shanghai, 201204, People's Republic of China

[d]Shanghai Advanced Research Institute, Chinese Academy of Sciences, Shanghai, 201210, People's Republic of China

Correspondence email: linhe@zjlab.org.cn; nzhang1@xjtu.edu.cn

**Funding information**    National Natural Science Foundation of China (grant No. 12161141012 to Nan Zhang; grant No. 51911530121 to Nan Zhang; grant No. U1732120 to He Lin; grant No. 21825102 to Jun Chen; grant No. 22075014 to Hui Liu).

**Synopsis** A general spherical harmonics method has been developed to transform X-ray total scattering data into anisotropic pair distribution functions. It provides an ideal tool to investigate the local structural change under the application of external stimuli.

**Abstract** A general spherical harmonics method is described for extracting anisotropic pair distribution functions (PDF) in this work. In the structural study of functional crystallized materials, the investigation of the local structures under the application of external stimuli, such as electric field and stress, is in urgent need. A well-established technique for local structure studies is PDF analysis, but the extraction of the X-ray PDF data is usually based on angular integrations of isotropic X-ray structure functions, which is no longer valid for the anisotropic responses of the materials under orientation-dependent stimuli. Therefore, we have developed an advanced spherical harmonics method to transform two-dimensional X-ray total scattering data into anisotropic PDF data, based on three-dimensional diffraction geometry and Fourier transform. The electrical-field-induced local structural change in the PbZr$_{0.54}$Ti$_{0.46}$O$_3$ ceramics is then presented to demonstrate the method's effectiveness.

## 1. Introduction

Total scattering and atomic pair distribution function (PDF) analysis techniques have been recognized as powerful tools for local structural investigations of many functional materials with local disordering (Proffen *et al.*, 2005; Choi *et al.*, 2014; Keen & Goodwin, 2015; Zobel *et al.*, 2015). The experimental PDFs can be obtained via a sine Fourier transform of the spherically averaged total-scattering structure function $S(Q)$. There are several well-developed program packages (Petkov, 1989; Qiu *et al.*, 2004; Soper & Barney, 2011; Juhás *et al.*, 2013, 2018) for neutron and X-ray total scattering data correction, treatment, and Fourier transform to produce the corresponding PDF data.

Because of the requirements of large $Q$-range and good signal-to-noise ratio at high-$Q$ region in total scattering experiments, early developments were mostly achieved in neutron scattering (Keen, 2020; Egami & Billinge, 2003). In recent decades, the Rapid-acquisition pair distribution function (RA-PDF) technique for X-ray scattering has been well developed and become very popular (Chupas *et al.*, 2003). In a typical RA-PDF experiment, total scattering data are collected with isotropic samples (powder, ceramic, glass, liquid etc.) using two-dimensional (2D) detectors, and the structure function $S(Q)$ are obtained by performing angular integration of the 2D scattering intensities. At the same time, efforts have been made to probe the responses of samples to external stimuli such as electric fields (Daniels *et al.*, 2010, 2009; Khansur, Rojac *et al.*, 2015) and stresses (Schader *et al.*, 2016; Daniel *et al.*, 2015, 2014). Especially for functional electronic materials like ferroelectrics and piezoelectrics, it is necessary to perform structural analysis on ceramic samples under the electric field. This is usually achieved by collecting the reciprocal space diffraction data of ceramics samples with the application of electric field in a transmission geometry, and analysing sectors of 2D diffraction data with different angles to the directions of the field (Pramanick *et al.*, 2011; Zhao *et al.*, 2018; Khansur, Hinterstein *et al.*, 2015; Wang *et al.*, 2014). The technical requirements, such as high-energy incident light (in order to penetrate the ceramics bulk) and a 2D detector, are very similar in total scattering experiments and *in-situ* electric-field diffraction experiments. It is therefore natural to consider combining them together to collect total scattering data and to study the local structural responses under the external stimuli. However, the current obstacle is that crystallized materials behave anisotropically after the application of electric field or stress, which is not applicable to the conventional PDF calculation methods assuming isotropic structure. On the theoretical front, Egami *et al.* (Egami *et al.*, 1995; Suzuki *et al.*, 1987; Dmowski *et al.*, 2010) employed spherical harmonics to investigate the bond-orientational anisotropy in metallic glasses under stress. Usher *et al*. (Usher *et al.*, 2015) extended the spherical harmonics approach to bulk polycrystalline dielectrics and ferroelectrics with electric field. In this total scattering experiment, the sample was required to be rotated by a certain angle towards the incident beam to minimize the angle between the electrical field and the average direction of the scattering vector.

In order to simplify the experimental operation and accurately process the anisotropic X-ray total scattering data, we have developed an advanced spherical harmonics-PDF method for the treatment of the 2D diffraction pattern. This modified spherical harmonics method implements spherical harmonics expansion and the least-squares method to yield anisotropic PDFs from $S(Q)$ obtained from the 2D total scattering pattern. We then apply it to produce anisotropic X-ray PDFs of a representative polycrystalline PbZr$_{0.54}$Ti$_{0.46}$O$_3$ sample to reveal the bond-length change induced by the electric field.

## 2. Theory

X-ray total scattering data is usually collected by a 2D flat-panel detector in transmission geometry. Fig. 1 illustrates the experimental setup and the Ewald sphere. The incident X-ray beam with the wavelength λ penetrates through the sample and is perpendicular to the applied stimuli. The reciprocal-space reference coordinate system is established in the way that **X**- and **Z**-axes are parallel to the beam and the stimuli, respectively. The sample is in the centre of the Ewald sphere with the radius of 2π/λ. The origin of the reference coordinate system is located at the intersection of the Ewald sphere and the direct beam. The relationship between the scattering vector ***Q*** and the coordinates of any chosen pixel on the detector is described by the scattering angle $2\theta$ and the azimuthal angle $\varphi$,

$$\boldsymbol{Q} = (Q \sin\theta, Q \cos\theta \sin\varphi, Q \cos\theta \cos\varphi) \tag{1}$$

where the modulus of the scattering vector, $Q$, is given by

$$Q = 4\pi \sin\theta / \lambda \tag{2}$$

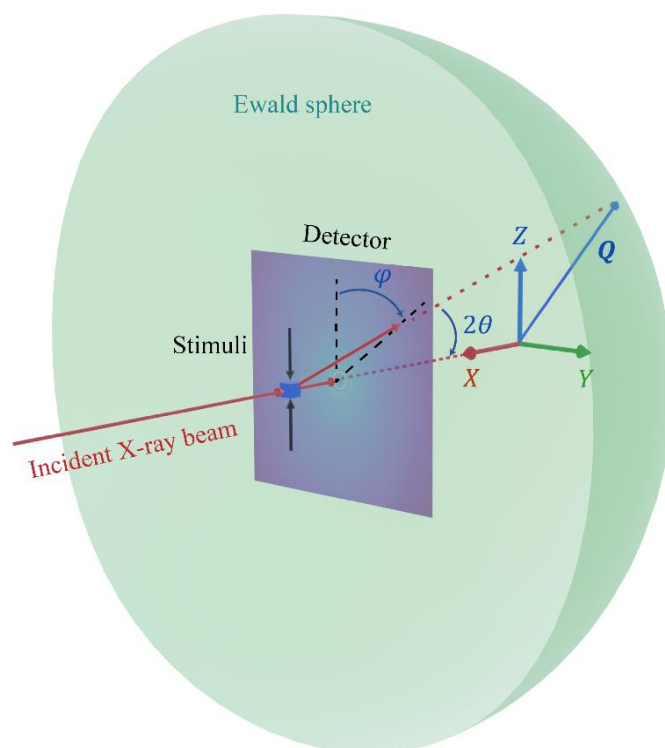

**Figure 1** The schematic for the X-ray total scattering experiment using a 2D detector. The incident X-ray beam penetrates through the sample along the **X**-axis with stimuli applied to the sample parallel to the **Z**-axis. The green sphere represents the Ewald sphere. The scattering angle $2\theta$ is the angle between the incident beam and the diffracted beam. The 2D diffraction pattern shows the Debye–Scherrer rings where the azimuthal angle $\varphi$ refers to the angle from the reference direction parallel to the Z-axis to a chosen line starting from the beam centre on the detector.

The three-dimensional (3D) PDF is calculated by 3D Fourier transform of the 3D total scattering data (Egami & Billinge, 2003; Harouna-Mayer *et al.*, 2022; Gong & Billinge, 2018),

$$\rho(\boldsymbol{r}) - \rho_0 = \frac{1}{8\pi^3} \int [S(\boldsymbol{Q}) - 1] \exp(i\boldsymbol{Q} \cdot \boldsymbol{r}) \mathrm{d}\boldsymbol{Q} \qquad (3)$$

where $\boldsymbol{r}$ is the atom pair vector, $\rho(\boldsymbol{r})$ is the atomic pair-density function and $\rho_0$ is the number density of the material. If the sample is isotropic, the modulus values of $S(\boldsymbol{Q})$ at different directions are mostly identical and can be directly integrated over a large range of $\boldsymbol{Q}/Q$. As a result, a spherical average over $S(\boldsymbol{Q})$ can be utilized to generate the isotropic PDF (Egami & Billinge, 2003),

$$G(r) = 4\pi r (\rho(r) - \rho_0) = \frac{2}{\pi} \int_0^\infty Q[S(Q) - 1] \sin(Qr) \, \mathrm{d}Q \qquad (4)$$

where $r$ is the modulus of $\boldsymbol{r}$.

In the anisotropic case, both the modulus and the direction $\boldsymbol{Q}/Q$ should be considered. The spherical harmonics expansion has been proven to be an effective approach to characterizing anisotropy (Dmowski *et al.*, 2010; Egami *et al.*, 1995; Suzuki *et al.*, 1987). Equation (3) indicates that $\rho(\boldsymbol{r})$- $\rho_0$ is the 3D inverse Fourier transform of $S(\mathbf{Q}) - 1$. We thus expend $\Delta\rho(\mathbf{r}) = \rho(\mathbf{r}) - \rho_0$ and $\Delta S(\mathbf{Q}) = S(\mathbf{Q}) - 1$ into spherical harmonics,

$$\Delta S(\mathbf{Q}) = \sum_{l=0}^{\infty} \sum_{m=-l}^{l} \Delta S_l^m(Q) Y_l^m(\mathbf{Q}/Q) \tag{5}$$

$$\Delta \rho(\mathbf{r}) = \sum_{l=0}^{\infty} \sum_{m=-l}^{l} \Delta \rho_l^m(r) Y_l^m(\mathbf{r}/r) \tag{6}$$

where $Y_l^m(\mathbf{x})$ is the orthonormalized spherical harmonics with degree $l$ and order $m$, $\mathbf{x}$ defines the direction of $\mathbf{Q}$ or $\mathbf{r}$. $\Delta S_l^m(Q)$ and $\Delta \rho_l^m(r)$ are the expansion components of $\Delta S(\mathbf{Q})$ and $\Delta \rho(\mathbf{r})$, respectively. In the case that $l = 0$ and $m = 0$, $\Delta S_l^m(Q)$ and $\Delta \rho_l^m(r)$ correspond to the isotropic functions. $\Delta S_l^m(Q)$ can be derived from $S(\mathbf{Q})$ by integrating over the solid angle as follows (Arfken *et al.*, 2013),

$$\Delta S_l^m(Q) = \int_\Omega S(\mathbf{Q}) Y_l^{m*}(\mathbf{Q}/Q) d\Omega \tag{7}$$

Where $\Omega$ is the solid angle and $Y_l^{m*}(\mathbf{x})$ is the complex conjugation of $Y_l^m(\mathbf{x})$.

Employing the Rayleigh expansion and the spherical harmonic addition theorem (Arfken *et al.*, 2013; Adkins, 2013), $\exp(i\mathbf{Q} \cdot \mathbf{r})$ can be expressed as:

$$\exp(i\mathbf{Q} \cdot \mathbf{r}) = 4\pi \sum_{l=0}^{\infty} \sum_{m=-l}^{l} i^l J_l(Qr) Y_l^m(\mathbf{Q}/Q) Y_l^{m*}(\mathbf{r}/r) \tag{8}$$

where $J_l(x)$ is the $l$th order spherical Bessel function. Substituting equation (5), (6), and equation (8) into equation (3), and using the orthonormality of spherical harmonics (Arfken *et al.*, 2013), the relationship between $\Delta S_l^m(Q)$ and $\Delta \rho_l^m(r)$ is then derived as,

$$\Delta \rho_l^m(r) = \frac{i^l}{2\pi^2} \int \Delta S_l^m(Q) J_l(Qr) Q^2 \mathrm{d}Q \tag{9}$$

$$\Delta S_l^m(Q) = 4\pi(-i)^l \int \Delta \rho_l^m(r) J_l(Qr) r^2 \mathrm{d}r \tag{10}$$

A typical calculation for $\Delta S_l^m(Q)$ is required to integrate $S(\mathbf{Q})$ over the entire solid angle as described in equation (7). A possible but brutal way is to collect the full scattering intensity in 3D reciprocal space and compute $\Delta S_l^m(Q)$ directly using equation (7). This total-scattering experiment can be performed by collecting diffraction frames during step-wise rotation of the sample, similar to an *in-situ* single-crystal diffraction experiment (Choe *et al.*, 2017; Gorfman *et al.*, 2020). The measurement, however, is time-consuming and requires sizeable operating memory (Egami & Billinge, 2003). On the other hand, certain symmetry exists in the 3D $S(\mathbf{Q})$ so that the experimental requirements and the computation process can be simplified in practical applications. First, if the stimuli are along the **Z**-axis, $S(\mathbf{Q})$ can be considered as being isotropic on the **XY**-plane. Also, because of the random grain distribution in ceramics, the diffraction intensities under the stimuli are symmetrical above and below the mirror plane perpendicular to **Z** passing through the incident beam. When these symmetry elements are introduced into equation (7), the $\Delta S_l^m(Q)$ values with $m \neq 0$ terms and odd $l$ terms are both zero (Egami *et al.*, 1995). Furthermore, with higher $l$ terms, $\Delta S_l^m(Q)$ values become substantially lower and may approach the noise level. These $\Delta S_l^m(Q)$ can be ignored if $l$ terms are higher than a given $l_{max}$ (typically $l_{max}$ = 2). Therefore, equation (5) can be simplified as,

$$\Delta S(\boldsymbol{Q}) = \sum_{l:\text{even}} \Delta S_l^0(Q) Y_l^0(\boldsymbol{Q}/Q) \approx \sum_{l:\text{even}}^{l_{max}} \Delta S_l^0(Q) Y_l^0(\boldsymbol{Q}/Q) \quad (11)$$

The anisotropic PDF along a specific direction $\boldsymbol{r}$ can be written as,

$$G(\boldsymbol{r}) = 4\pi r \Delta \rho(\boldsymbol{r}) = 4\pi r \left( \sum_{l:\text{even}}^{l_{max}} \Delta \rho_l^m(r) Y_l^m(\boldsymbol{r}/r) \right) \quad (12)$$

It is worth noting that $l = 0$ corresponds to the isotropic component in the anisotropic PDF. Equation (12) is identical to equation (4) for $l_{max}$ = 0 and can be used to extract the isotropic PDF.

## 3. Application

To verify the reliability of our spherical harmonic approach, an anisotropic X-ray total scattering dataset was processed to generate the anisotropic PDF data. The *in-situ* X-ray total scattering experiment was performed on PbZr$_{0.54}$Ti$_{0.46}$O$_3$ ceramics sample under the electric field. The data were collected at the 11-ID-C beamline of the Advanced Photon Source (APS) using an incident beam with the wavelength of 0.11714 Å and a PerkinElmer flat-panel detector. The distance between the sample and the detector is 35.46 cm. The applied electric field is 5 kV/mm, denoted by $\boldsymbol{E}$. More details of the experiment can be found in Liu *et al.* (Liu *et al.*, 2022). As shown in Fig. 2 (a), the 2D diffraction pattern was divided into 18 $\varphi$-azimuthal sectors from 0° to 90°, and each sector was integrated into one-dimensional diffraction data using Fit2D (Hammersley, 2016). Structure functions dependent on different $Q$ scalar at different $\varphi$ (denoted as $S'(Q)$ in this work) were normalized and converted from the raw diffraction intensities using PDFgetX2 (Qiu *et al.*, 2004).

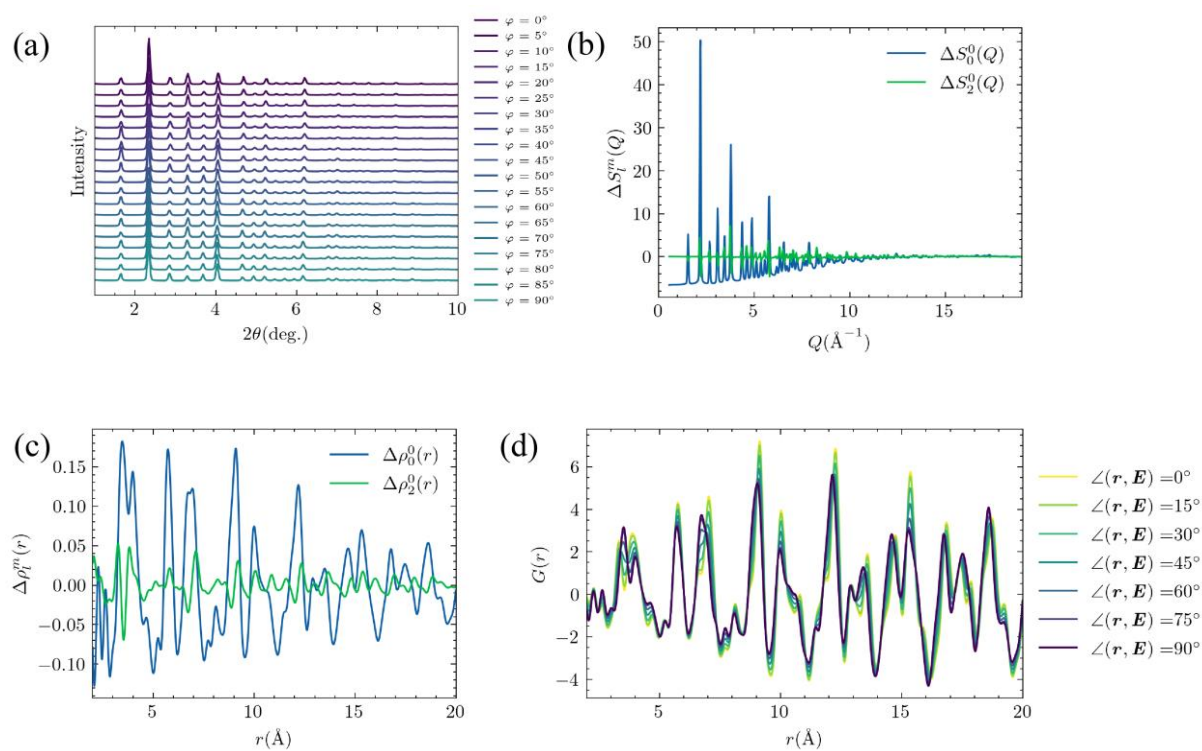

**Figure 2** (a) The 1D intensity profiles integrated from $\varphi$-azimuthal sectors from 0° to 90° in the 2D diffraction pattern of PbZr$_{0.54}$Ti$_{0.46}$O$_3$ ceramics. (b) The calculated zero- and second-degree expansion components $\Delta S_l^0(Q)$. (c) The calculated zero- and second-degree expansion components $\Delta \rho_l^0(r)$. (d) The anisotropic PDFs with different angles between $r$ and $E$ from 0° to 90°. $\angle(r, E)$ refer to the angle between $r$ and $E$.

For each azimuthal sector, $\varphi$ is usually considered to be almost equal to the angle between scattering vectors and applied stimuli since the sample-detector distance is rather small in most of the total-scattering experiments. However, equation (1) indicates that these two angles are actually never exactly the same for any $\varphi$ angle. To avoid the possible errors introduced by this approximation, it is necessary to use the 3D coordinates of all vectors involved in the process. $S'(Q_i)$ versus $Q_i$ data produced by PDFgetX2 from an azimuthal sector $i$ are thus converted to $S(\boldsymbol{Q}_i)$ versus $\boldsymbol{Q}_i$ data using equation (1) and (2). The $\Delta S(\boldsymbol{Q}_i)$ data from different sectors, which have equal $Q$ magnitude but different $\boldsymbol{Q}_i/Q$ direction, represents $\Delta S(\boldsymbol{Q})$ anisotropy distribution along different directions. Following equation (11), these $\Delta S(\boldsymbol{Q}_i)$ can be expanded into a series of spherical harmonics equations in order to obtain expansion components $\Delta S_l^0(Q)$ as follows,

$$\begin{cases} \Delta S(\boldsymbol{Q}_1) = \sum_{l:\text{even}}^{l_{max}} \Delta S_l^0(Q) Y_l^0(\boldsymbol{Q_1}/Q) \\ \Delta S(\boldsymbol{Q}_2) = \sum_{l:\text{even}}^{l_{max}} \Delta S_l^0(Q) Y_l^0(\boldsymbol{Q_2}/Q) \\ \quad\quad\quad\quad \vdots \\ \Delta S(\boldsymbol{Q}_i) = \sum_{l:\text{even}}^{l_{max}} \Delta S_l^0(Q) Y_l^0(\boldsymbol{Q_i}/Q) \end{cases} \quad (13)$$

Under the typical presumption that $l_{max}$ = 2, the $\Delta S_l^0(Q)$ values are obtained by solving the above equations using the least-squares method, while the expansion components $\Delta\rho_l^0(r)$ are calculated by performing the integrations in equation (9). The calculated $\Delta S_l^0(Q)$ and $\Delta\rho_l^0(r)$ when $l$ = 0, 2 are shown in Fig. 2(b) and (c). Once $r$ is determined, $\Delta\rho(r)$ can be calculated by performing the summation in equation (6) and then be utilized to obtain the isotropic PDF and anisotropic PDFs using equation (12). Fig. 2(d) illustrates the anisotropic PDFs with various angles between $r$ and $E$ denoted as ∠($r$, $E$). The whole process of extracting the angular dependent anisotropic PDFs is summarized in the flow chart in Fig. 3.

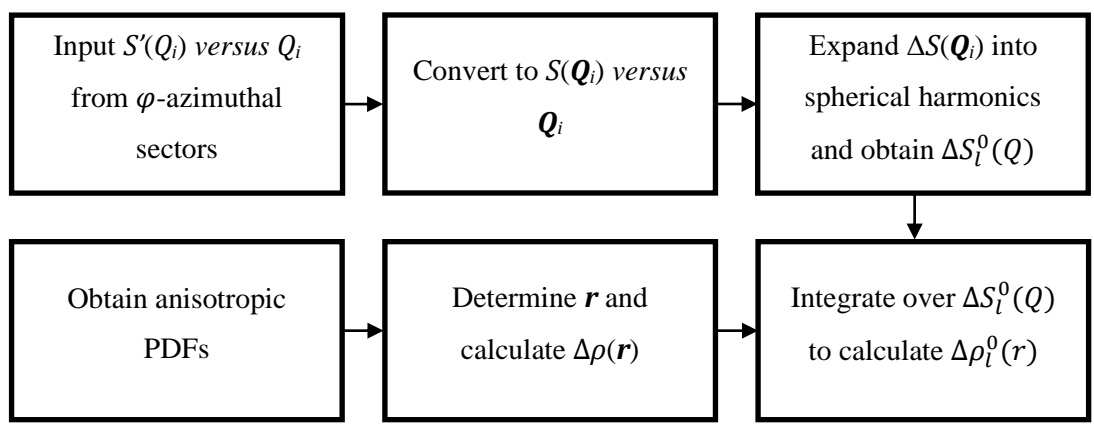

**Figure 3** The flow-chart for processing anisotropic PDFs using the modified spherical harmonics method.

Previous structural refinement (Zhang *et al.*, 2018) of the isotropic PDF suggests that the local structure of PbZr$_{0.54}$Ti$_{0.46}$O$_3$ has monoclinic symmetry under zero electric field. The majority of the Pb displacement directions are on the M$_A$ mirror plane but still close to the ⟨111⟩ direction. Anisotropic PDF patterns shown in Fig. 2(d) exhibit that the peak positions and intensities differ significantly with $E$, indicating distinct anisotropic local structural behaviours.

Considering the fact that the X-ray PDF peaks involving oxygen atoms have very low intensities, only Pb-B (B = Zr/Ti) and Pb-Pb/B-B pairs within a unit-cell scale are considered in our preliminary analysis. As illustrated in Fig. 4(a), Pb-B bonds are divided into short, medium and long bonds in monoclinic symmetry (Zhao *et al.*, 2017), denoted by $R_s$, $R_m$ and $R_l$. Pb-Pb/B-B pairs can be categorized into three groups: the nearest-neighbour pairs along the ***r*** //⟨100⟩ direction, the next-nearest-neighbour pairs along the unit-cell face diagonal (***r*** //⟨110⟩), and the pairs along the unit-cell body diagonal (***r*** //⟨111⟩). Two representative anisotropic PDFs with ***r*** //***E*** and ***r*** ⊥***E*** and the isotropic PDF are shown in Fig. 5, in which the correspondence between PDF peaks and these atom pairs is identified. Among them, ***r*** //⟨111⟩ peaks obviously split into two peaks with slightly different *r* values.

The first Pb-B pair has the *r* value between 3.3 and 3.8 Å, and is along one of the eight ⟨111⟩ directions, which is of a small angle with the local polarization direction in an $M_A$ unit cell. Each anisotropic PDF pattern at this range is composed of the Pb-B length distributions in the specific grains or domains in which the angle between the body diagonal of the unit cell and the electric field is the same as ∠(***r***, ***E***), and one possible domain orientation for the $G($***r*** //***E***$)$ configuration is shown in Fig. 4(b). It is observed in the $G($***r*** //***E***$)$ pattern in Fig. 5 that the position of the $R_s$ pair becomes even smaller and the $R_l$ pair becomes even longer than those in the isotropic PDF. On the other hand, in the $G($***r*** ⊥***E***$)$ pattern, the $R_s$ and the $R_l$ pairs tend to move closer to the $R_m$ position, forming an almost single-peak shape with the average bond length similar to the original $R_m$. This can be explained by the fact that for the domains and grains having the polarizations almost parallel to ***E***, the Pb atoms tend to move forward under the field, which decreases the distance between Pb and the nearest B cation and increases the distance between Pb and the farthest B cation. While in the grains that ⟨111⟩ is perpendicular to ***E***, Pb displacements are prohibited, making the 8 Pb-B bonds have similar lengths. A similar trend is also observed in the Pb-Pb / B-B pairs along ⟨111⟩ in the range of *r* = 6.5 ~ 7.4 Å that the peak positions show obvious shift for ***r*** //***E*** state while remain generally fixed for ***r*** ⊥***E*** state, indicating an increased distortion when the polarization has a small angle with the electric field. Moreover, there is a drastic intensity exchange of the ***r*** //⟨111⟩ peaks for the ***r*** //***E*** and ***r*** ⊥***E*** patterns, which indicates that there are more domains with polarization directions almost parallel to the electric field formed in the ***r*** //***E*** grains while less of them appearing in the ***r*** ⊥***E*** grains. This clearly shows domain switching behaviour after the application of electrical field, accompanied by electric-field-induced lattice distortion. This preliminary analysis of the PZT ceramics demonstrates that the anisotropic PDFs produced using our method effectively reveal the field-induced local structural behaviours of ferroelectric perovskites.

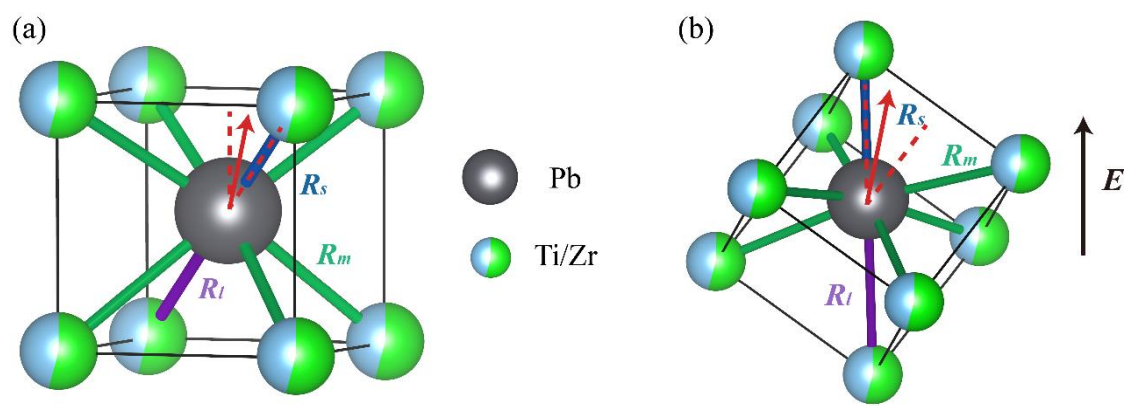

**Figure 4** A monoclinic PbZr$_{0.54}$Ti$_{0.46}$O$_3$ model with space group *Cm*. (b) A possible domain orientation in the PbZr$_{0.54}$Ti$_{0.46}$O$_3$ under the electric field. The red arrow indicates the direction of the polarization vector rotating on the (110) plane. The oxygen atoms in the model are not shown. Blue, green and purple bonds mark short ($R_s$), medium ($R_m$) and long ($R_l$) Pb-B bonds, respectively.

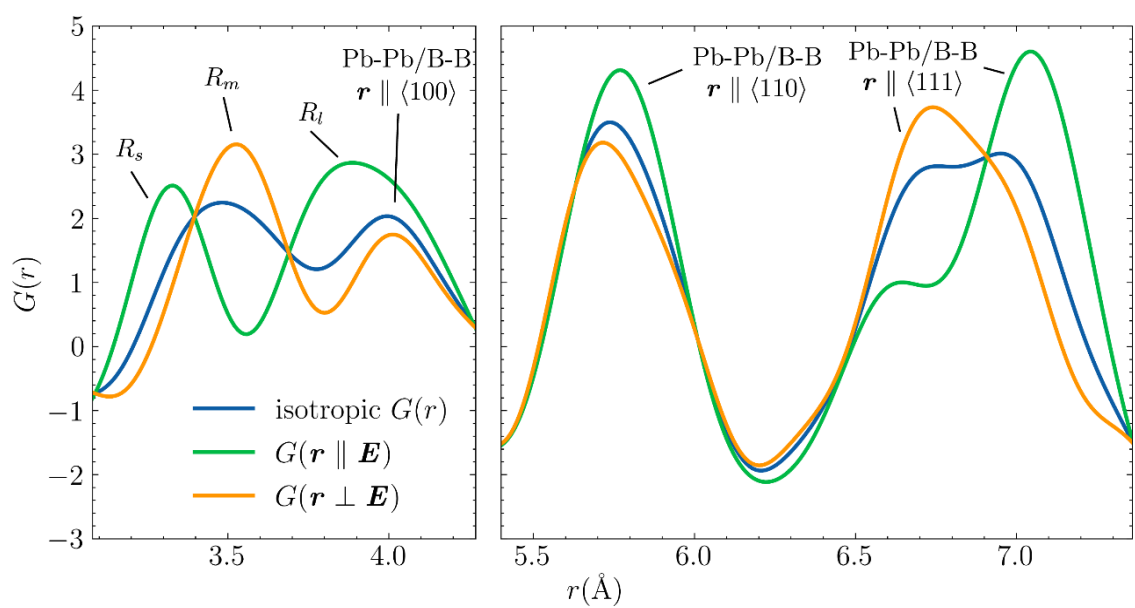

**Figure 5** The isotropic PDF (blue), anisotropic PDFs $G(r /\!/ E)$ (green) and $G(r \perp E)$ (yellow) showing Pb-B and Pb-Pb/B-B bonds in PbZr$_{0.54}$Ti$_{0.46}$O$_3$ in the very short *r*-range.

### 4. Program

We generated a graphical user interface (GUI) software, AniPDF, based on the spherical harmonic method described above to process anisotropic total scattering data interactively. The software is written in Python and operates on Linux and Windows platforms. AniPDF performs analysis operations to transform $S'(Q)$ into anisotropic PDFs using Python package NumPy (Harris *et al.*, 2020), SciPy (Virtanen *et al.*, 2020), and Matplotlib (Hunter, 2007). The GUI is constructed using wxpython, a cross-platform GUI toolkit. AniPDF can be employed together with other process programs such as Fit2D and PDFgetX2 to obtain anisotropic PDFs from 2D diffraction patterns.

The GUI shown in Fig. 6 has three major components: data panel, parameter panel, and output panel. The upper data panel is used to load raw one-dimensional total-scattering data. The input files must be in the $Q, S'(Q)$ two-column format and be collected under the same experimental conditions. The $\varphi$ angle corresponding to each input file is specified by the user. In the middle parameter panel, the user can adjust the X-ray wavelength and parameters related to spherical harmonics. The direction vector of the atom pair $r$ should be given in the Cartesian coordinates or the spherical coordinates. In the bottom output panel, the user can also set the output folder and choose the type of output file. By pressing the "Run" button, the entire computation process procedure is launched and anisotropic PDFs are then exported as text files.

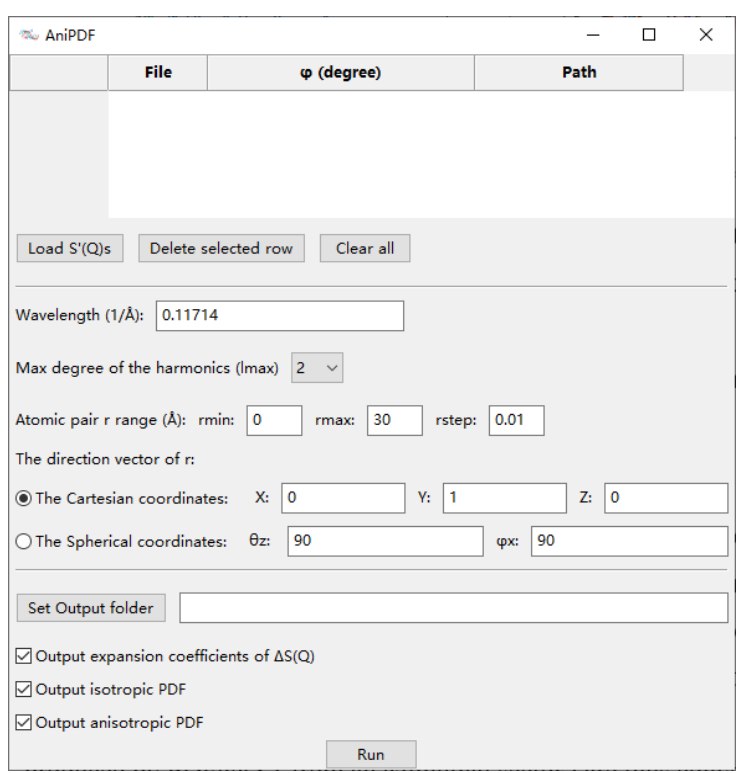

**Figure 6** The AniPDF graphical user interface consisting of the upper data panel, the middle parameter panel and the bottom output panel.

## 5. Conclusion and Outlook

We have introduced a modified spherical harmonics method for processing reliable anisotropic PDFs for the samples under the external stimuli. This method gives insight into the local structural behaviour and aids in the understanding of the stimuli-response mechanism. For ferroelectrics and piezoelectrics, it is a straight forward tool for probing domain switching and lattice distortion, which is widely believed to be the origin of high piezoelectric response. The method has been successfully applied to the PZT sample under the external electrical field. It should be noted that the anisotropy introduced in this example has a unique axis, and the atom distribution shows axial symmetry. However, the spherical harmonics method can be extended for anisotropic conditions without a unique axis simply by considering $\Delta S_l^m(Q)$ with $m \neq 0$ terms and odd $l$ terms. Therefore, our method can also be applied to the local structural studies of more intricate material systems and complicate experimental settings to fulfil various types of structural responses in functional materials.

**Acknowledgements**     We thank Xiangyun Qiu for the assistance with the data processing. The authors would like to thank 11-ID-C beamline of the Advanced Photon Source (APS) and BL02U2 beamline of Shanghai Synchrotron Radiation Facility (SSRF).